\documentclass[doublecol]{epl2}
\usepackage{graphicx}
\usepackage[utf8]{inputenc}
\usepackage{amsmath}
\usepackage{verbatim}
\usepackage{amssymb}
\usepackage{listings}
\usepackage{color}
\usepackage{xifthen}

\newcommand{\eqtwo}[2]{eqs~(\ref{#1}) and~(\ref{#2})}
\newcommand\subfig[2]{{Fig.~\ref{#1}{#2}}}
\newcommand\subcap[1]{{(#1)}}
\newcommand\bigO[1]{\ensuremath{\mathcal{O}({#1})}}

\newcommand\unifb[1]{\text{unif}\glb #1 \grb}

\newcommand\unifd[1]{\text{unif}\gld #1 \grd}

\newcommand\NChains{\ensuremath{n}}
\newcommand\NMinChains{\ensuremath{m}}

\newcommand\TVD{\ensuremath{\text{TVD}}}
\newcommand\TVDMATH[1]{\|#1\|_\text{TV}}

\newcommand\Lfree{\ensuremath{L_{\text{free}}}}

\newcommand\fsma{forward swap Metropolis algorithm}
\newcommand\prob[1][]{%
  \ifthenelse{\isempty{#1}}%
    {\pi}% if #1 is empty
    {\pi^{(#1)}}% if #1 is not empty
}
\newcommand{\pitilde}{\tilde{\pi}}
\newcommand\probtilde[1][]{%
  \ifthenelse{\isempty{#1}}%
    {\pitilde}% if #1 is empty
    {\pitilde^{(#1)}}% if #1 is not empty
}

\newcommand{\half}{\frac{1}{2}}
\newcommand\diff[1]{\mathrm{d}#1}
\newcommand{\mean}[1]{\left\langle #1 \right\rangle}
\newcommand{\glb}{\left(}  % ' group left b' 
\newcommand{\grb}{\right)}  % ' group right b' 
\newcommand{\glc}{\left[}  % ' group left c'
\newcommand{\grc}{\right]}  % ' group right c' 
\newcommand{\gld}{\left\{}  % ' group left d' 
\newcommand{\grd}{\right\}}  % ' group right d' 
\newcommand{\VEC}[1]{\mathbf{#1}}
\newcommand{\SET}[1]{\{#1\}}
\newcommand{\TO}{,\ldots,}
\newcommand{\xvec}{\VEC{x}}

\newcommand{\expa}[1]{\mathrm{e}^{#1}}   % high exponential groupings a
\newcommand{\expb}[1]{\exp \glb #1 \grb} % low exponential with groupings b
\newcommand{\expc}[1]{\exp \glc #1 \grc} % low exponential with groupings c
\newcommand{\lifted}[2]{\ensuremath{(#1,#2)}}
\newcommand{\ACAL}{\mathcal{A}}
\newcommand{\RCAL}{\mathcal{R}}  %  mathcal
\newcommand{\FCAL}{\mathcal{F}}  %  mathcal
\newcommand{\NCAL}{\mathcal{N}}  %  mathcal
\newcommand{\quot}[1]{``#1''}
\newcommand{\PLUSPLUS}{+ \dots +}
\newcommand{\sinb}[2][]{\sin^{#1} \glb #2 \grb}  % sin-brace,  with - ()
\newcommand{\cosb}[2][]{\cos^{#1} \glb #2 \grb}  % cos-brace,  with - ()
\newcommand{\cosc}[2][]{\cos^{#1} \glc #2 \grc}  % cos-brace,  with - []

\newcommand{\eq}[1]{eq.~(\ref{#1})}
\newcommand{\fig}[1]{Fig.~\ref{#1}}
\begin{document}
\date{\today}\title{Mixing and perfect sampling in one-dimensional particle 
systems}
\author{Ze Lei\inst{1,2}\thanks{\email{ze.lei@ens.fr}} \and Werner Krauth 
\inst{1,2}
\thanks{\email{werner.krauth@ens.fr}}}
\institute{
\inst{1} Laboratoire de Physique Statistique, D\'{e}partement de physique
de l'ENS, Ecole Normale Sup\'{e}rieure, PSL Research University, Universit\'{e}
Paris Diderot, Sorbonne Paris Cit\'{e}, Sorbonne Universit\'{e}s, UPMC
Univ. Paris 06, CNRS, 75005 Paris, France \\
\inst{2} Max-Planck-Institut f\"{u}r Physik komplexer Systeme, N\"{o}thnitzer 
Str. 38, 01187 Dresden, Germany
}
\shortauthor{Z. Lei \etal}

\abstract{ We study the approach to equilibrium of the event-chain Monte
Carlo (ECMC) algorithm for the one-dimensional hard-sphere model. Using the
connection to the coupon-collector problem, we prove that a specific version
of this local irreversible Markov chain realizes perfect sampling in \bigO{N^2
\log N} events, whereas the reversible local Metropolis algorithm requires
\bigO{N^3 \log N} time steps for mixing. This confirms a special case of an earlier
conjecture about \bigO{N^2 \log N} scaling of mixing times of ECMC and of the
forward Metropolis algorithm, its discretized variant. We furthermore prove
that sequential ECMC (with swaps) realizes perfect sampling
in \bigO{N^2} events. Numerical simulations indicate a cross-over towards
\bigO{N^2 \log N} mixing for the sequential \fsma,
that we introduce here. We point out open mathematical questions and
possible applications of our findings to higher-dimensional statistical-physics
models.}

\maketitle
%\tableofcontents

\section{Sampling, mixing, perfect sampling, stopping rules}
Ever since the 1950s\cite{Metropolis1953}, Markov-chain Monte Carlo (MCMC)
methods have ranked among the most versatile approaches in scientific
computing. Monte Carlo algorithms strive to sample a probability distribution
$\prob$. For an $N$-particle system in statistical mechanics, with particle
$i\in \SET{1 \TO N}$ described by coordinates $x_i$, sampling $\pi$ corresponds
to generating configurations $\xvec = \SET{x_1 \TO x_N}$ distributed with the
Boltzmann probability $\prob(\xvec) \propto \expc{-\beta E(\xvec)}$, where $E$
is the system energy and $\beta$ the inverse temperature. For problems  where
$\xvec$ lies in a high-dimensional discrete or continuous space $\Omega$, this
sampling problem can usually not be solved directly \cite{Devroye1986,SMAC}.

MCMC consists instead in sampling a probability distribution $\prob[t]$
that evolves with a time $t$.  The initial probability distribution, at time
$t=0$, $\prob[t=0]$, can be sampled directly.  Often, it is simply composed
of a single configuration, so that $\prob[0]$ is a $\delta$-function  on an 
explicitly given
configuration $\SET{x_1 \TO x_N}$. In the limit $t
\to \infty$, the distribution $\prob[t]$ evolves from the initial one towards
the target probability distribution $\prob = \lim_{t \to \infty} \prob[t]$.
Besides the development of MCMC algorithms that approach the limit distribution
$\prob$ as quickly as possible for any initial distribution $\prob[0]$,
a key challenge in MCMC consists in estimating the time scale on which
the time-dependent distribution $\prob[t]$, which depends on $\prob[0]$, is
sufficiently close to $\prob$ that the two agree for all intents and purposes.
This program has met with considerable success in some models of statistical
physics, for example for the local Glauber dynamics in the two-dimensional
Ising model \cite{Martinelli1999,Lubetzky2012}.

The difference between two (normalized) probability distributions $\prob$
and $\probtilde$ is often quantified by the total variation distance (TVD)
\cite{Levin2008,Diaconis2011},
\begin{align}
 \TVDMATH{\probtilde-\prob} &=  \half   \int_{\Omega} |\probtilde(\xvec) - \prob(\xvec)| \diff \xvec 
\label{equ:FirstDefinitionTVD}\\
  &= \max_{\ACAL \subseteq \Omega} |\probtilde(\ACAL)- \prob(\ACAL)|,
  \label{equ:SecondDefinitionTVD}
\end{align}
which satisfies $0 \le \TVDMATH{\probtilde-\prob} \le 1$. The mixing time, the 
most relevant figure
of merit for a Monte Carlo algorithm, is defined as the time $t$ after which
the \TVD\ (with $\probtilde \equiv \prob[t]$ in \eq{equ:SecondDefinitionTVD})
is smaller than a given threshold $\epsilon$, for any initial distribution
$\prob[0]$.  Although it is of great conceptual importance, the TVD cannot
usually be computed. In statistical physics, this is already because
the normalization of the Boltzmann weight, the partition function
$Z = \int_{\Omega} \expb{-\beta E}$, is most often unknown. Also, the
distribution $\prob[t](\xvec)$ is not known explicitly.
It is because
of this difficulty that practical simulations often carry systematic uncertainties
that are difficult to quantify, and that heuristic convergence criteria for  the
approach towards equilibrium in MCMC
abound\cite{Berg2004book,LandauBinderBook2013,SMAC}. They most often involve
time-correlation functions of observables, rather than the probability
distribution itself (as in
\eqtwo{equ:FirstDefinitionTVD}{equ:SecondDefinitionTVD}).

In rare cases, MCMC algorithms allow for the definition of a stopping rule
(based on the concept of a strong stationary time\cite{Levin2008}),
that yields a simulation-dependent value of $t$ at which the configuration
is sampled \emph{exactly} from the distribution $\prob$.  The value of $t$
now often depends on the realization of the Markov chain (that is, on the
individual sampled moves and, ultimately, on the drawn random numbers). 
The framework of stopping rules can be used to bound the mixing
time\cite{Levin2008}. Stopping rules
exist for  quite intricate models, as for example the Ising model, using the
mixing-from-the-past framework\cite{ProppWilson1996,SMAC}.

The great majority of Markov-chain Monte Carlo algorithms are reversible
(they satisfy the detailed-balance condition). This is the case for
example for all algorithms that are based on the Metropolis or the heat-bath
algorithms\cite{Metropolis1953,SMAC}, which allow reversible MCMC algorithms to
be readily constructed for any distribution $\prob$, that is, for an arbitrary
energy $E(\xvec)$.  In recent years, however, irreversible MCMC
methods based on the global balance condition have shown considerable
promise\cite{Turitsyn2011,FernandesWeigelCPC2011,Bernard2009,Michel2014JCP,
KapferKrauth2017}. In these algorithms, $\prob[t]$ approaches
$\prob$ for long times, but the flows no longer vanish. One
particular irreversible Markov chain, the event-chain Monte
Carlo (ECMC) algorithm\cite{Bernard2009,Michel2014JCP},
has proven useful for systems ranging from hard-sphere
models\cite{Bernard2011} to spin systems\cite{LeiKrauthXY2018},
polymers\cite{KampmannKierfeldJCP2015,Harland2017} and
to long-range interacting ensembles of molecules, such as
water\cite{FaulknerAllAtom2018}, where the Coulomb interaction plays a dominant
role\cite{KapferKrauth2016}. Although there have been many indications of the
algorithm's power, no exact results were known for the mixing behavior of ECMC,
except for the case of a single particle, $N=1$ \cite{Diaconis2000}.

In the present paper, we rigorously establish ECMC mixing times and
stopping rules of the model of $N$ hard spheres on a one-dimensional
line with periodic boundary conditions (a circle). Reversible
MCMC algorithms for this model and its variants were analyzed
rigorously\cite{RandallWinklerInterval2005,RandallWinklerCircle2005} and
irreversible MCMC algorithm were discussed in detail\cite{KapferKrauth2017}.
The 1D hard-sphere model and reversible and irreversible MCMC algorithms
are closely related to the symmetric exclusion process (SEP) on a periodic
lattice\cite{Lacoin_2017_SSEP} and to the totally asymmetric simple exclusion
process (TASEP) \cite{GwaSpohnPRL1992,ChouTASEP2011,BaikLiu2016}. For
ECMC, an algorithm that is closely related to the lifted Metropolis
algorithm\cite{KapferKrauth2017}, we compute the $\TVD$ in a special case,
and obtain the mixing times as a function of the parameter $\epsilon$.
We confirm the $\bigO{N^2 \log N}$ mixing time that had been conjectured on
the basis of numerical simulations\cite{KapferKrauth2017}. Furthermore, we
obtain a stopping rule for ECMC. We moreover present sequential variants of the
forward Metropolis algorithm and the ECMC algorithm. For the latter, we prove
an \bigO{N^2} exact-sampling result that seems however not to generalize the
discretized version of the algorithm.

\section{Hard spheres in 1D, reversible Monte Carlo}

The mixing and convergence behavior of Markov chains for particle
systems has been much studied. As for hard spheres in 2D and
above, phase transitions have only been identified by numerical
simulation\cite{HooverRee1968,Alder1962,Bernard2011}, it is natural that few
rigorous results are available for the convergence and mixing behavior of MCMC
algorithms in $D>1$\cite{Wilson2000,Kannanrapidmixing2003}. We thus restrict
our attention to the 1D hard-sphere model with periodic boundary conditions and
treat both the discrete and the continuous cases.

\begin{figure}[htb]
\includegraphics[width = \linewidth]{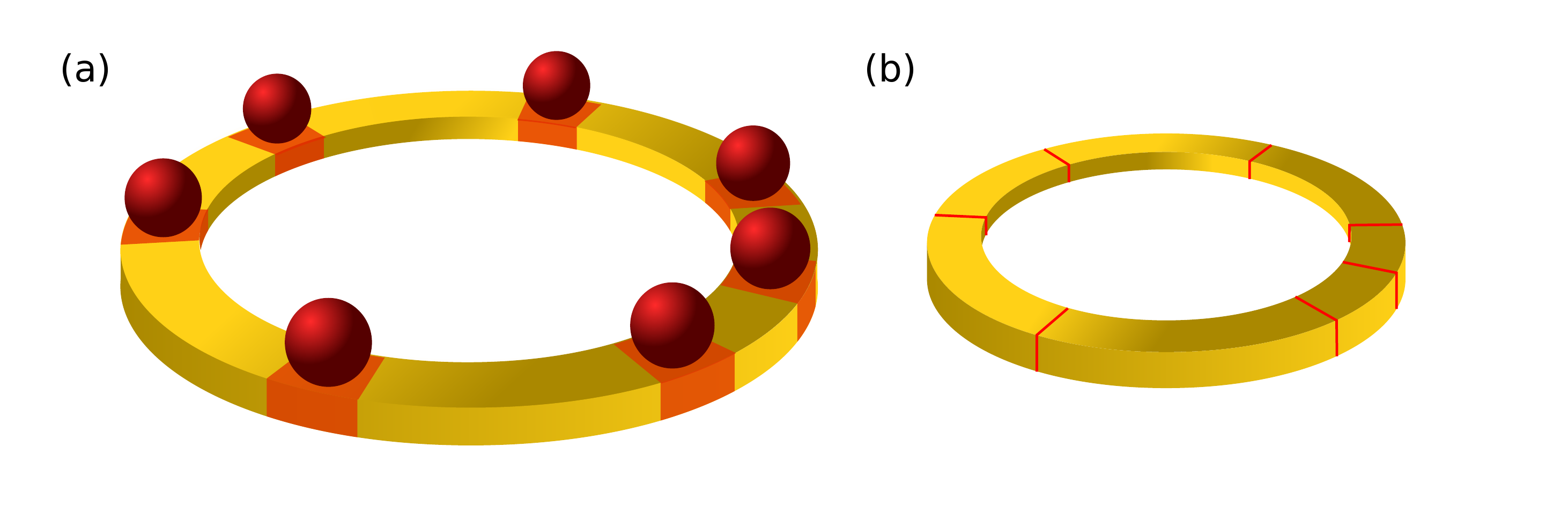}
\caption{1D hard-sphere model with periodic boundary conditions. 
\subcap{a}: $N$ spheres of diameter $d$ on a ring of length $L$.
\subcap{b}:  $N$ point particles on a ring of length $\Lfree = L -  N d$. 
Configurations and local MCMC algorithms are equivalent for both 
representations. } 
\label{fig:OneDParticlesPoints}
\end{figure}

The 1D hard-sphere model can be represented as $N$ spheres of diameter $d$
on a line of length $L$ with periodic boundary conditions (that is, on a ring, 
see \subfig{fig:OneDParticlesPoints}{a}). A valid configuration $a$ of $N$ 
spheres has unit statistical weight $\pi(a)=1$. Spheres do not overlap, so that the 
distance between sphere centers, and in particular between neighboring 
spheres, is larger than $d$. Each valid configuration of $N$  hard spheres is 
equivalent to a configuration of $N$ point particles on a ring of length $\Lfree 
= L - N d > 0$ (see \subfig{fig:OneDParticlesPoints}{b}), and the partition
function of the model equals $S = \Lfree^N$, which proves the absence of a phase
transition.

We only consider local Markov chains, where the move of sphere $i$ is accepted
or rejected based solely on the position of its neighbors.  One might think
that this requires the distribution $p(\epsilon)$ to vanish for $|\epsilon| >
2 d$. We rather implement locality by rejecting a move of sphere $i$ not only
if the displacement leads to an overlap, but also if sphere $i$ would hop over
one of its neighbors. In this way, any local Monte Carlo move of spheres on a
circle corresponds to an equivalent move in the point-particle representation
(for which there are no overlaps and moves are rejected only because they
represent a hop over a neighbor). The dynamics of both models is the same. This
implies that the MCMC dynamics of the 1D hard-sphere model has only a trivial
density dependence.

Although we will study Markov chains that relabel spheres, 
we are interested only in the relaxation of quantities that can be expressed
through the unlabeled distances between neigboring spheres. This excludes the
mixing in permutation space of labels or the self-correlation of a given sphere with
itself (or another labeled sphere) at different times. Trivial uniform rotations are
thus also neglected.

Detailed balance consists in requiring:
\begin{equation}
 \prob(a) p(a \to b) = \prob(b) p(b \to a), 
 \label{equ:DetailedBalance}
\end{equation}
where $p(a \to b)$ is the conditional probability to move from configuration  
$a$ to configuration $b$. 
The heat-bath algorithm is a local reversible MCMC algorithm. At each time
step, it replaces a sampled sphere $i$ randomly in between
its neighbors. The heat-bath algorithm mixes in
at least \bigO{N^3} and at most \bigO{N^3\log N} time steps\cite{RandallWinklerCircle2005},
although numerical simulations clearly favor the latter possibility (\bigO{N^3
\log N})\cite{KapferKrauth2017}.\footnote{As mentioned, we do not consider 
uniform rotations of the total system, which would mix only on a time
scale \bigO{N^4}.} For the one-dimensional hard-sphere model on a line without
periodic boundary condition, the mixing time \bigO{N^3\log N}
is rigorously proven\cite{RandallWinklerInterval2005}.

Analogous to the heat-bath algorithm, the reversible Metropolis algorithm
also satisfies the  detailed-balance condition: At each time step, a randomly 
chosen 
sphere $i$ attempts a move by $\epsilon$ taken from some probability
distribution. The move is rejected if the proposed displacement
$\epsilon$ is larger than the free space in the direction of the proposed
move ($x_{i_+} - x_i -d$ for $\epsilon >0$) or behind it ($x_i - x_{i_-}
-d$ for $\epsilon <0$) (where we suppose that $i_+$ is the right-hand-side 
neighbor of $i$, etc, and imply periodic boundary conditions).
In the point-particle model, the
equivalent move is rejected if the particle would hop over one or more
of its neighbors and is accepted otherwise. Rigorous results for mixing
times are unknown for the Metropolis algorithm, but numerical
simulations clearly identify \bigO{N^3 \log N} mixing as for the heat-bath
algorithm\cite{KapferKrauth2017}. In the discrete 1D hard-sphere model on the
circle with $L$ sites and $N$ particles, the Metropolis algorithm is implemented
in the socalled simple exclusion process (SEP), where at each time step, a
randomly chosen particle attempts to move with equal probability to each of its
two adjacent sites. The move is rejected if that site is already occupied. The
mixing time of the SEP is $\sim(4\pi^2)^{-1}N L^2\log N $ (for $L \geq 2 N$)
\cite{Lacoin_2017_SSEP}.

\section{From the forward Metropolis to the event-chain algorithm}

\begin{figure}[htb]
\begin{center}
\includegraphics[width = 0.8\linewidth]{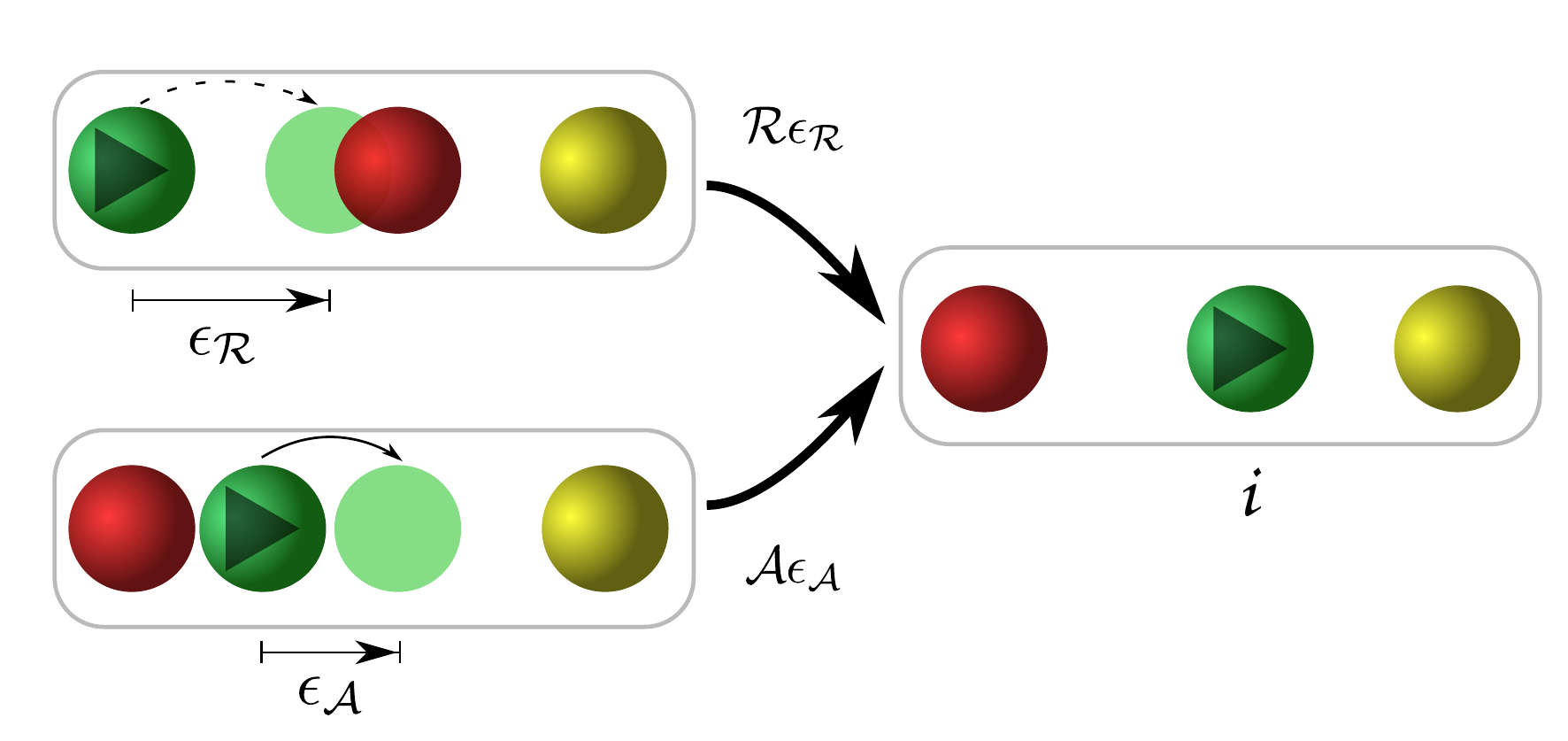}
\end{center}
%\vspace*{-5mm}
\caption{Flow of the \fsma\  into a
configuration $\lifted{a}{i}$ (the active sphere $i$ is shown in green). A
rejected sphere move, by a displacement $\epsilon_\RCAL$ (upper case), entails
a swap and contributes
$\RCAL_{\epsilon_\RCAL}$. An accepted sphere move, by a displacement
$\epsilon_{\ACAL}$ (lower case), contributes $\ACAL_{\epsilon_\ACAL}$. For any
$\epsilon$, one of the flows equals one, and the other zero
[$(\RCAL_{\epsilon}, \ACAL_{\epsilon} ) \in \SET{(0,1),(1,0)} $].}
\label{fig:MoveSwap}
\end{figure}

Irreversible Monte Carlo algorithms violate the detailed-balance condition
of \eq{equ:DetailedBalance} but instead satisfy the weaker global-balance
condition
\begin{equation}\label{eq:weak_global_balance}
 \sum_b \prob(b) p(b \to a) = \prob(a).
\end{equation}
Together with the easily satisfiable aperiodicity and irreducibility
conditions\cite{Levin2008}, the global-balance condition ensures that
the steady-state solution corresponds to the probability $\prob$,
but without necessarily cancelling the net flow $ \prob(a) p(a \to b)
-  \prob(b) p(n \to a)$ between configurations $a$ and $b$ (\emph{cf}
\eq{equ:DetailedBalance}). Here, we take up the forward Metropolis algorithm
studied earlier, in a new variant that involves swaps. This allows us to arrive
at an exact mixing result.

In the \fsma\footnote{The forward Metropolis algorithm introduced
earlier\cite{KapferKrauth2017} did not feature swaps.}, at each time step, a
randomly chosen sphere $i$ attempts to move by a random displacement $\epsilon$
with a predefined sign (that for clarity, we take to be positive). If the
move is rejected (that is, the displacement $\epsilon$ does not yield a
valid hard-sphere configuration), the sphere swaps its label with the sphere
responsible for the rejection (see the upper move in \fig{fig:MoveSwap}). Else,
if the displacement $\epsilon$ is accepted, the sphere $i$  simply moves
forward (see the lower move in \fig{fig:MoveSwap}).  The total flow into a
configuration $\lifted{a}{i}$ (that is, the $N$-sphere configuration $a$ with
the active sphere $i$) is:
\begin{equation}
    \FCAL(a,i)  = \int_0^\infty \diff \epsilon p(\epsilon) 
    \underbrace{\glc \ACAL_\epsilon(a,i) + \RCAL_\epsilon(a,i) \grc}_{=1\ 
\text{(see \fig{fig:MoveSwap})}}= 1 = \pi(a), 
\end{equation}
so that the algorithm satisfies global balance. The swap allows both the
rejected and the accepted moves into the configuration $\lifted{a}{i}$
to involve the sphere $i$ only. The forward swap Metropolis algorithm is
equivalent (up to relabeling) to the forward Metropolis algorithm treated
earlier if at each time step the active sphere $i$ is sampled randomly.
The mixing time of this algorithm was conjectured to be \bigO{N^{5/2}},
based on numerical simulations\cite{KapferKrauth2017}. This agrees with the
proven mixing time scale of the totally asymmetric simple exclusion process
(TASEP)\cite{BaikLiu2016}.

\begin{figure}[htb]
\includegraphics[width = \linewidth]{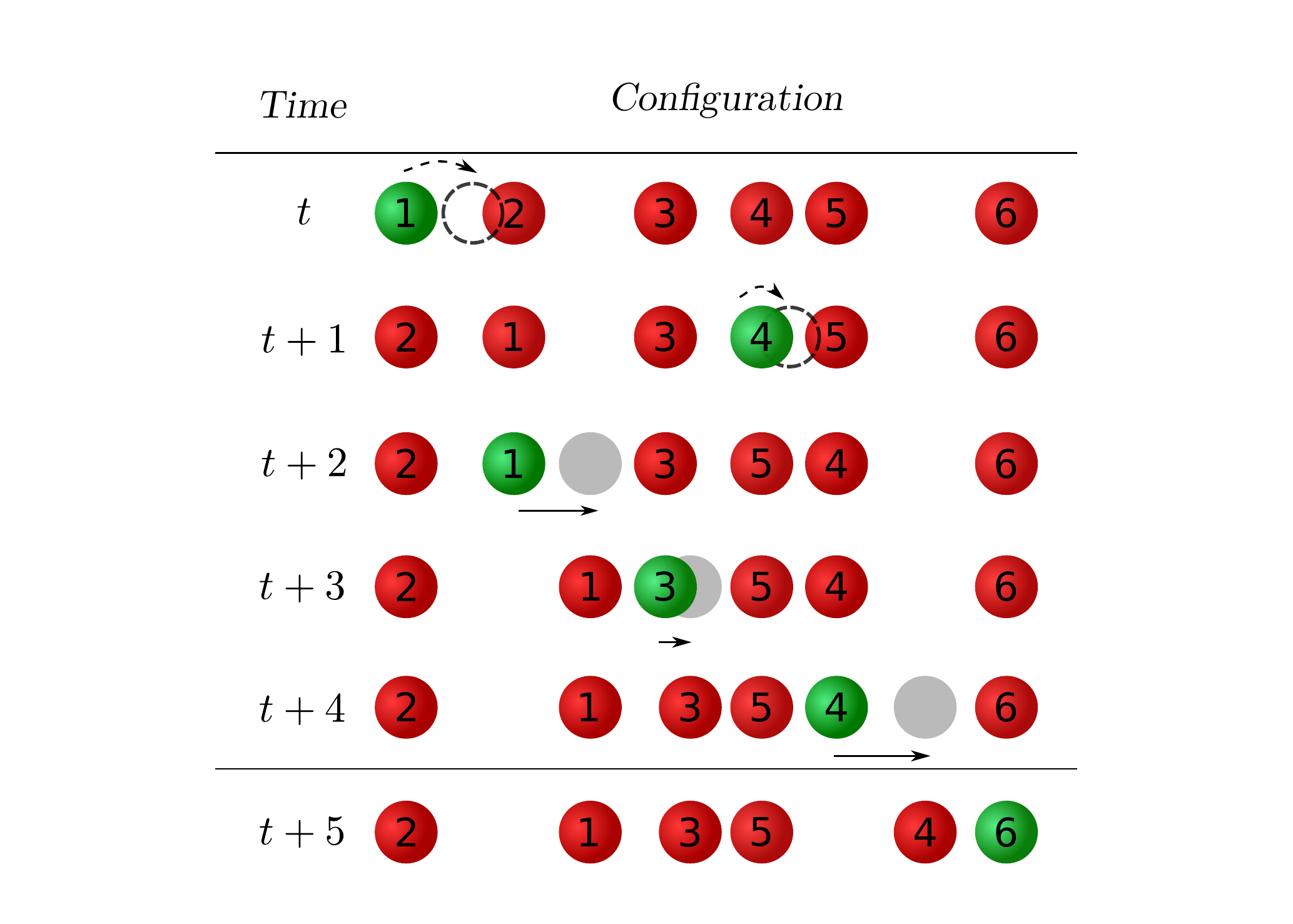}
\caption{Forward swap Metropolis algorithm, with configurations
$\xvec_t \TO \xvec_{t+5}$.  The active sphere is sampled randomly at each time
step so that the swaps have no other action than to relabel the spheres.}
\label{fig:ForwardMetropolis}
\end{figure}

The \fsma\ satisfies global balance for any choice of the sphere $i$ and any
step-size distribution $p(\epsilon)$. This implies that the active-sphere
index $i$ need not be sampled randomly for the algorithm to remain valid.
This distinguishes it from the forward Metropolis algorithm (without the swaps)
treated in previous work\cite{KapferKrauth2017}. In particular, the sphere $i$
can be active for several chains in a row.  The algorithm, run with the 
following sequence of active spheres:
\begin{equation}
\dots,
\underbrace{i,i,\dots,i, i}_{\text{chain $\NChains$}},
\underbrace{j,j,\dots,j, j}_{\text{chain $\NChains +1$}},
\underbrace{k,k,\dots,k, k}_{\text{chain $\NChains + 2$}},
\dots, 
\label{equ:LiftedForwardPartSwap}
\end{equation}
is equivalent to the lifted forward Metropolis algorithm studied
earlier\cite{KapferKrauth2017}, if the active spheres $i,j,k,\dots$
in \eq{equ:LiftedForwardPartSwap} are sampled randomly.  The algorithm
naturally satisfies the global balance condition, and again, each individual
move attempts a displacement by a distance $\epsilon$ sampled from a given
distribution $p(\epsilon)$ that vanishes for negative $\epsilon$, and the
chain lengths (number of repetitions of $i,j,k,\dots$) $\NChains, \NChains
+ 1, \dots$ are sampled from a distribution.  Numerical simulations have
lead to the conjecture that this algorithm mixes in \bigO{N^2 \log N} time
steps\cite{KapferKrauth2017}.

ECMC is the continuous-time limit of the lifted forward Metropolis algorithm,
with the simultaneous limits $\epsilon \to 0$ and $l \to \infty$, but $\glb
\mean{\epsilon} l \grb \to \ell$, where the chain length $\ell$ on the scale
$\Lfree$, is again sampled from a given probability distribution. In the
point-particle representation of \subfig{fig:OneDParticlesPoints}{(b)}, one
\quot{event} chain of the ECMC algorithm simply moves the active particle $i$
from its initial position $x_i$ to $x_i + \ell$. It advances
the time as $t \to t + \ell$, and increments the number of chains as
$\NChains \to \NChains + 1$. The number of eponymous \quot{events} of ECMC
(the number of changes of the active sphere) then grows approximately as
$\sim (\ell / \Lfree) N$. When $\ell \sim \unifb{0,\Lfree}$, this places
particle $i$ at a random position on a ring. For this special uniform
distribution of chain lengths, a perfect sample is clearly obtained once all
particles were at least once picked as the active particle. This
situation will now be analyzed in terms of the coupon-collector problem (see
\cite{ErdosRenyi_Coupon_1961,Blom1994}).

\begin{figure}[htb]
\begin{center}
\includegraphics[width = 0.7 \linewidth]{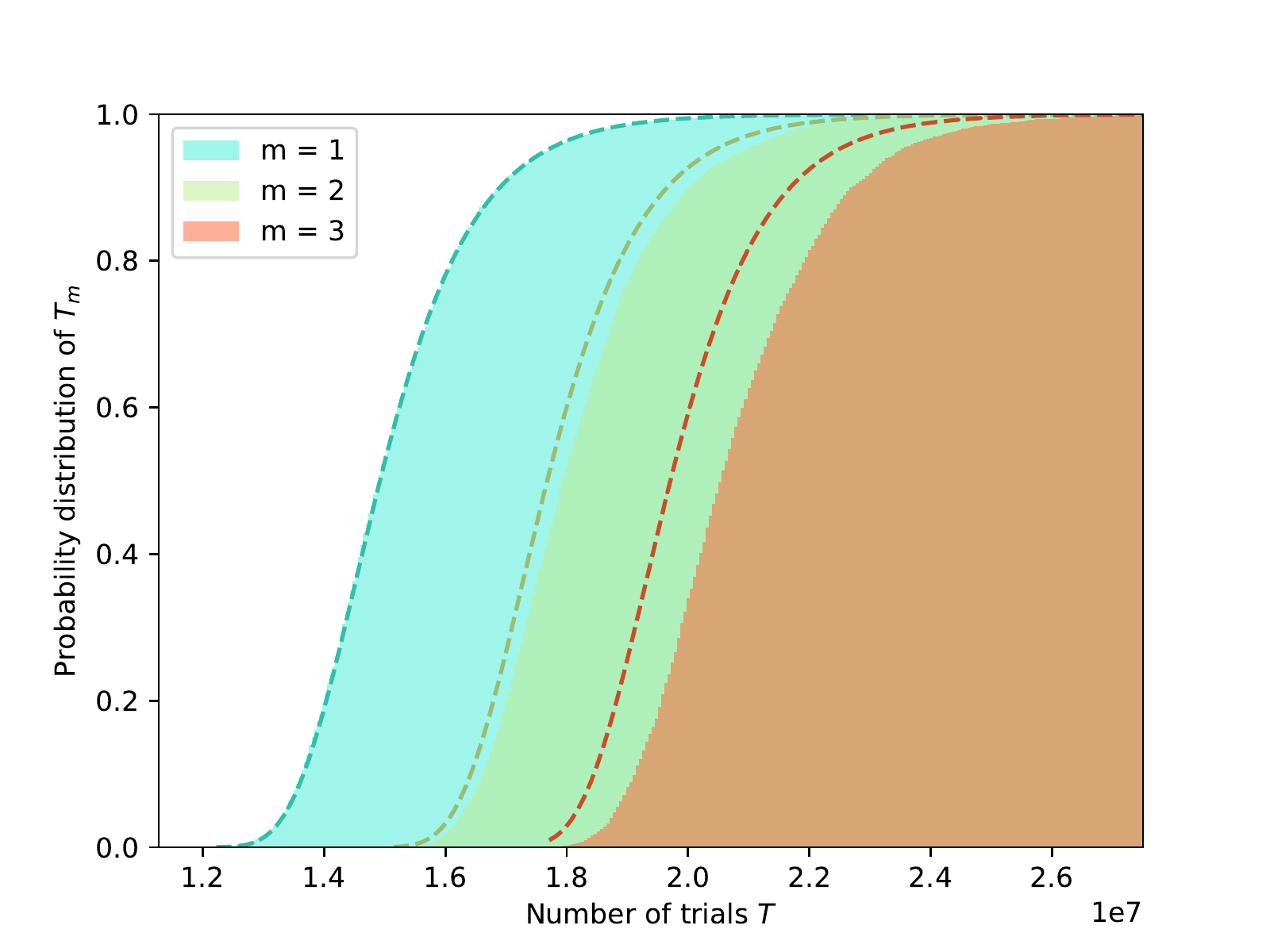} \end{center} 
\caption{Cumulative probability of the coupon collector problem ($m=1$), and of 
the $m$-coupon 
collector problem for $m=2 $ and $m=3$. 
Numerical simulations for $N = 2^{20}$ particles are compared to the asymptotic 
tail probability of \eq{equ:MCoupons}.} 
\label{fig:CouponCollector}
\end{figure}

For the ECMC with $\ell \sim \unifb{0, \Lfree}$, the TVD can be expressed by the
probability that at least one particle has never been picked as an active 
particle of a chain. Without restriction, we suppose that the initial 
configuration is the compact state $\xvec = \SET{0,0 \TO 0}$. We also measure
time in the number of chains $\NChains$ 
($\NChains$ translates into an MCMC time as  $t(n) = \mean{\ell} n$ and is 
easily converted into the number of events).
In \eq{equ:SecondDefinitionTVD}, the set $\ACAL$ is 
\begin{equation}
 \ACAL = \SET{\xvec\ |\ \exists\ i\ \text{with}\ x_i = 0}.
\end{equation}
Also, clearly, $\prob[n](\ACAL)$ equals the probability that at
least one particle has never been picked as an active particle of a chain, 
whereas $\prob(\ACAL)=0$, as it is a lower-dimensional subset of $\Omega$.
From \eqtwo{equ:FirstDefinitionTVD}{equ:SecondDefinitionTVD}, therefore
(for $N \to \infty$):
\begin{equation}
\TVDMATH{ \prob[n] - \prob} \sim 1 -
\expc{-\expb{-\frac{n - N \log N}{N}}},
\label{equ:TVDCoupon1} 
\end{equation}
where we used the analytically known asymptotic tail probability for the 
coupon-collector problem\cite{ErdosRenyi_Coupon_1961}
(see \fig{fig:CouponCollector}).

Rather than computing the difference between $\prob[n]$ and $\prob$ at a fixed
number $n$ of chains, one can simply run ECMC (with $\ell \sim 
\unifb{0,\Lfree}$) until the time at which chains with any of the $N$ 
particles as active ones have completed. The expected number of chains or, in 
the language of the coupon-collector problem, the expected value of $\NChains_1$ 
to \quot{collect the last coupon} is given by
\begin{equation}
 \mean{\NChains_1}  = N H_N = N \log N + \gamma N + \half +  \bigO{1/N},
\label{equ:OneCoupon} 
\end{equation}
where $H_N = \frac11 + \frac12  \PLUSPLUS \frac1N$ is the $N$th harmonic
number and $\gamma = 0.5772...$ is the Euler-Mascheroni constant.
The distribution of this number of chains can be obtained from the tail 
distribution contained in \eq{equ:TVDCoupon1}
(see \fig{fig:CouponCollector}). In both ways, we see that mixing takes
place after \bigO{N \log N} chains (corresponding to \bigO{N^2 \log N}
events), confirming, for a special distribution of $\ell$,  an earlier
conjecture\cite{KapferKrauth2017}. The discussed mixing
behavior of ECMC can more generally be obtained for distributions $\ell \sim 
\unifb{c, c + \Lfree}$ with arbitrary (and even with negative) $c$.  In our 
special case, choosing $c = - \Lfree/2$ would lead to the smallest number of 
individual events. In view of the practical applications of ECMC, it appears 
important to understand whether this dependence on the distribution of $\ell$ 
rather than on its mean value has some relevance for the simulation of discrete 
1D models, and whether it survives in higher dimensions, and for continuous 
(non-hard-sphere) potentials.

We next consider more general distributions, namely the uniform distribution 
$\ell \sim \unifb{0,\lambda \Lfree}$, as well as the Gaussian
distribution $\NCAL(\mu, \sigma^2)$, where $\mu$ is the mean value and 
$\sigma$ the standard deviation. Again, particles
are effectively independent and we conjecture the mixing time (which can now never lead
to perfect sampling) to be governed by the particle which has moved the least
number, $m$,  of times. This is equivalent to the $m$-coupon generalization of 
the coupon-collector problem\cite{ErdosRenyi_Coupon_1961}, 
whose tail probability is given by:
\begin{equation}
 P(\NChains_m) = \expb{-\Upsilon/ (m-1)!}\\
\end{equation} 
with
\begin{equation}
 \Upsilon =
  \expc{-\frac{\NChains_m - N \log N - (m-1) N \log \log N}{N}}
\label{equ:MCoupons} 
\end{equation}
(see \fig{fig:CouponCollector}). This means that the number of chains to collect 
each of the $N$ coupons at least $m$  times only add an $N \log \log N$ 
correction to the general $N \log N$ scale of chains. 
 
To gain intuition, we now compute the TVD for the single-particle problem (for 
which $\ell \equiv \epsilon$). 
For simplicity, we set $\Lfree = 1$ (measure the 
standard deviations in units of $\Lfree$). The TVD for  
chain lengths $\ell_i \sim 
\unifb{0,\lambda}$, 
as discussed, equals the one for  $\ell_i \sim 
\unifb{- \lambda/2,\lambda/2}$.  
The sum of $\NMinChains$ chains then follows the 
distribution: 
\begin{equation}
p^{\text{unif}}_\NMinChains(x)  = \int_{-\infty}^{\infty } \diff 
t 
\expa{-2\pi i tx} \glc \frac{\sinb{\pi\lambda t}}{\pi\lambda 
t}\grc^{\NMinChains}
\end{equation}

\begin{figure}[htb]
\includegraphics[width = \linewidth]{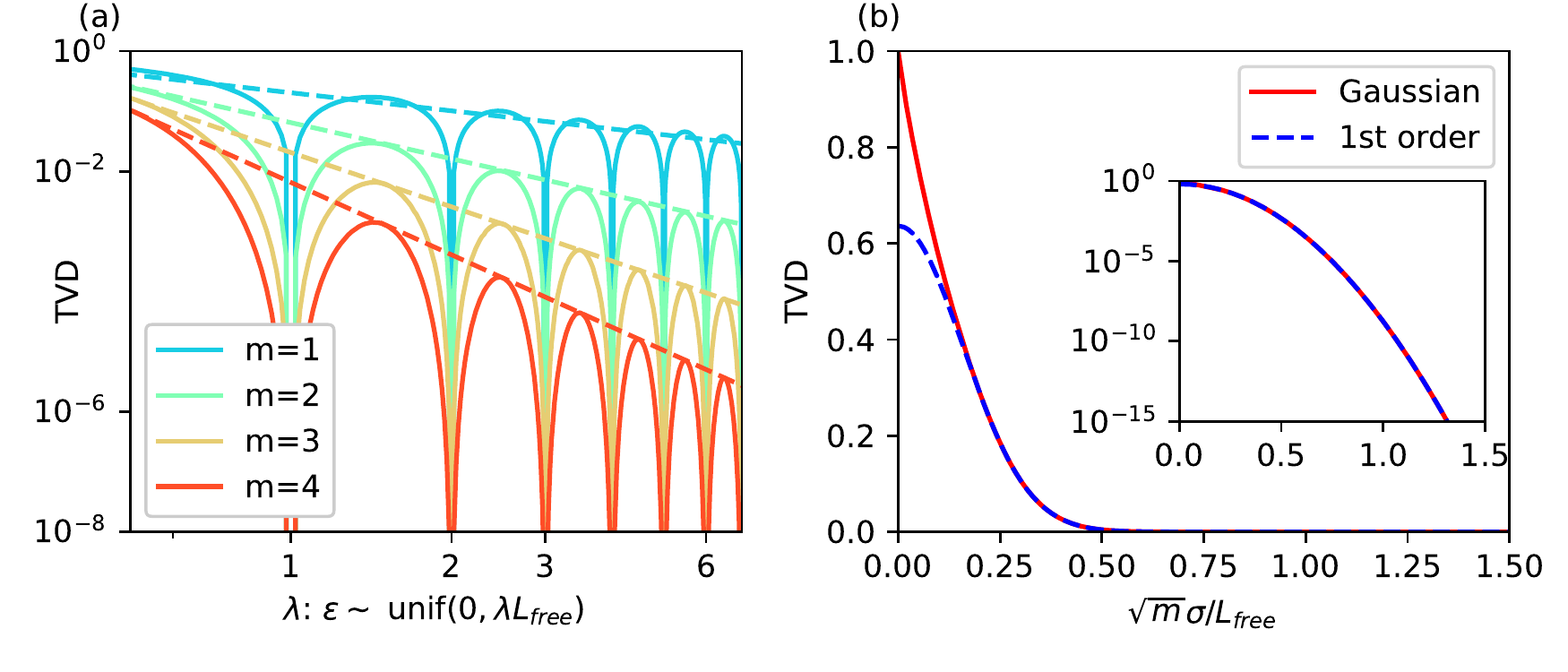}
\caption{
TVD for a single sphere on a ring with uniform and Gaussian
distributions of $\ell \equiv \epsilon$.  \subcap{a}: TVD after $m$ 
displacements
$\epsilon \sim \unifb{0, \lambda \Lfree}$. The TVD trivially vanishes 
for integer $\lambda$. 
Peaks decay as  $\frac{2}{\pi} (\pi \lambda)^{-m}$ (for $m \to \infty$). 
\subcap{b}:
TVD for $m$ Gaussian displacements with standard deviation $\sigma$, compared 
with its first order
approximation from the Jacobi $\vartheta$ function
(see \eq{equ:TVDGaussian}).
The inset illustrates the good agreement of the approximation on
a logarithmic scale.}
\label{fig:TVD}
\end{figure}

Using the Poisson summation formula and subtracting 
the equilibrium distribution $\prob = 1$, we find:
\begin{equation*}
\sum_{k=-\infty} ^{\infty} p^{\text{unif}}_\NMinChains(x + k) -1 = \sum_{k 
\in \mathbb{N^+}} 2\glc \frac{\sinb{\pi k\lambda}}{\pi k\lambda} 
\grc^\NMinChains 
\cos{\glb 2\pi k x\grb} .
\end{equation*}
The total variation distance for chain lengths $\ell_i \sim \unifb{0, 
\lambda}$ thus satisfies:
\begin{multline}
 \TVDMATH{ \prob[m] - \prob}   \\ = \int_0^1 dx \left|
 \sum_{k 
\in \mathbb{N^+}} \glc \frac{\sinb{\pi k\lambda}}{\pi k\lambda} 
\grc^\NMinChains 
\cos{\glb 2\pi k x\grb } 
 \right |   \\
 \sim
\frac {2}{\pi}\left| \frac{\sinb{\pi \lambda}}{\pi \lambda} \right|^\NMinChains 
 \  
(\text{for} 
\quad 
\NMinChains \to 
\infty).
\label{equ:TVDSingleUniCosSum}
\end{multline}
The TVD trivially vanishes for integer $\lambda$ (see \fig{fig:TVD}{a}).
Its peaks decay as $\frac{2}{\pi} (\pi\lambda)^{-\NMinChains}$.

For Gaussian-distributed chain lengths $\ell_i \sim \NCAL(\mu, 
\sigma^2)$, the sum of $\NMinChains$ chains is distributed as:
\begin{equation}
 \sum_{i=1}^{\NMinChains} \ell_i \sim \mathcal{N}(\NMinChains \mu, 
\NMinChains \sigma^2).
\label{equ:SumOfGaussians}
\end{equation}
With $\vartheta_3$ the Jacobi theta function, we now have
\begin{multline}
\sum_{k=-\infty} ^{\infty} p^{\text{Gauss}}_m(x + k) - 1=  \vartheta_3\glc \pi 
(x+ 
\mu), \expb{- 2 \pi^2 \NMinChains \sigma ^2} \grc  \\ =  2 \sum_{k=1}^{\infty} 
\expb{- 2 k^2 
\pi^2 
\NMinChains \sigma^2} \cosc{2 k \pi (x+\NMinChains\mu)}.
\end{multline}
The total variation distance for the distribution of \eq{equ:SumOfGaussians} 
satisfies:
\begin{multline}
 \TVDMATH{ \prob[m] - \prob} \\ = \int_0^1 dx \left|
 \sum_{k=1}^{\infty} \expb{- 2 k^2 \pi^2 \NMinChains \sigma^2 } \cosb{2 k \pi 
x}
 \right | \\
 \sim
\frac{2}{\pi} \expb{- 2 \pi^2 \NMinChains \sigma^2} \  (\text{for} \quad 
\NMinChains \sigma^2 \to 
\infty)
\label{equ:TVDGaussian}
\end{multline}
(see \subfig{fig:TVD}{b}).

Both for the uniform and the Gaussian distribution, the single-sphere 
$\TVD$ decreases exponentially with the number $m$ of displacements (which are 
equivalent to single-particle chains). We expect the same behavior for the 
$N$-sphere problem, where $m$ is now the number of chains for the $m$-coupon 
problem. 

\section{Sequential forward Metropolis, sequential ECMC}

\begin{figure}[htb]
\includegraphics[width = \linewidth]{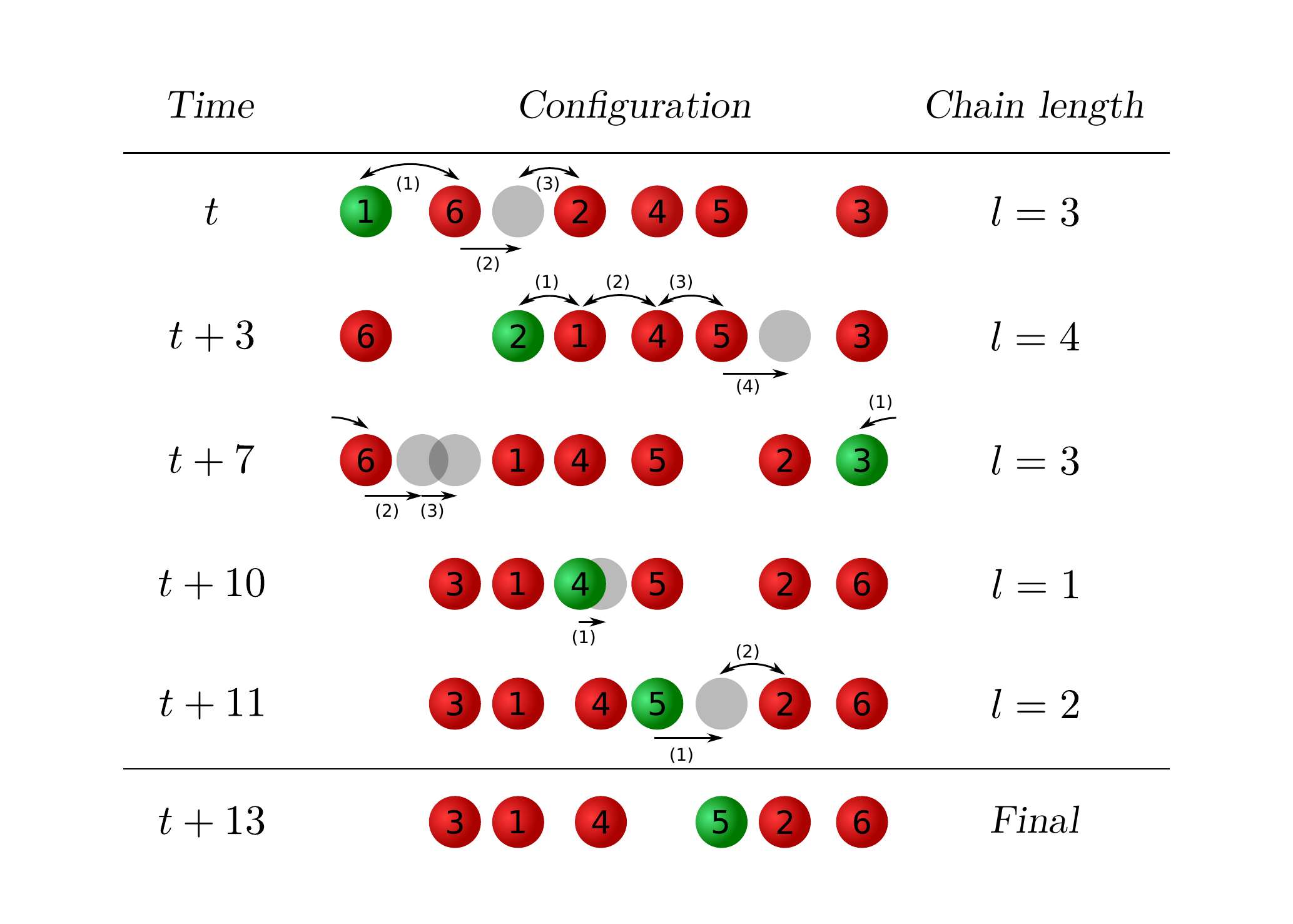}
\caption{Sequential lifted forward Metropolis algorithm (with swaps).
Configurations $\xvec_t \TO \xvec_{t+13}$ sampled through five chains with
active sphere $1,2 \TO 5$ are shown. Chain lengths are $l_1 = 3 \TO l_5 = 2$.
Each sphere displacement $\epsilon_t > 0$ is either accepted or, if rejected,
it induces a swap, so that the same sphere remains active throughout a chain.}
\label{fig:ForwardMetropolisIllustration}
\end{figure}
ECMC, with randomly sampled initial spheres and a standard deviation of the
chain-length distribution  $\sigma \sim \Lfree$, mixes in \bigO{N^2 \log N}
events (corresponding to \bigO{N \log N} chains). In the label-switching
framework of ECMC, each chain consists in advancing the particle $i$ by a
distance $\ell$
times, and both the ECMC and the forward-Metropolis versions are correct.
Instead of sampling the active sphere for each chain, so that the 
coupon-collector aspect necessarily brings in the $\log N$ factor in the
scaling of  mixing times, we may
also sequentially increment the active-sphere index for each chain (see
\fig{fig:ForwardMetropolisIllustration}): 
\begin{equation}
\dots,
\underbrace{i,\dots, i}_{\text{chain}\ i},
\underbrace{i+1,\dots, i+1}_{\text{chain}\ i+1},
\underbrace{i+2,\dots, i+2}_{\text{chain}\ i+2},
\dots, 
\label{equ:LiftedForwardPartSwapSeq}
\end{equation}
(where particle numbers are implied modulo $N$). Sequential ECMC, with a 
distribution $\ell_i \sim \unifb{0,\Lfree}$ produces an exact sample in 
\bigO{N^2} events (corresponding to exactly $N$ chains).

Evidently, the analysis of \eqtwo{equ:TVDSingleUniCosSum}{equ:TVDGaussian} can 
be applied to the sequential ECMC
with distributions such as $\unifb{0,\lambda\Lfree}$ and,  more generally,
distributions with $\sigma \sim \Lfree$.
After each \quot{sweep} of chains, the TVD factorizes, and we expect mixing to 
take place after \bigO{N} chains (corresponding to \bigO{N^2} events).

\begin{figure}[htb]
\includegraphics[width = \linewidth]{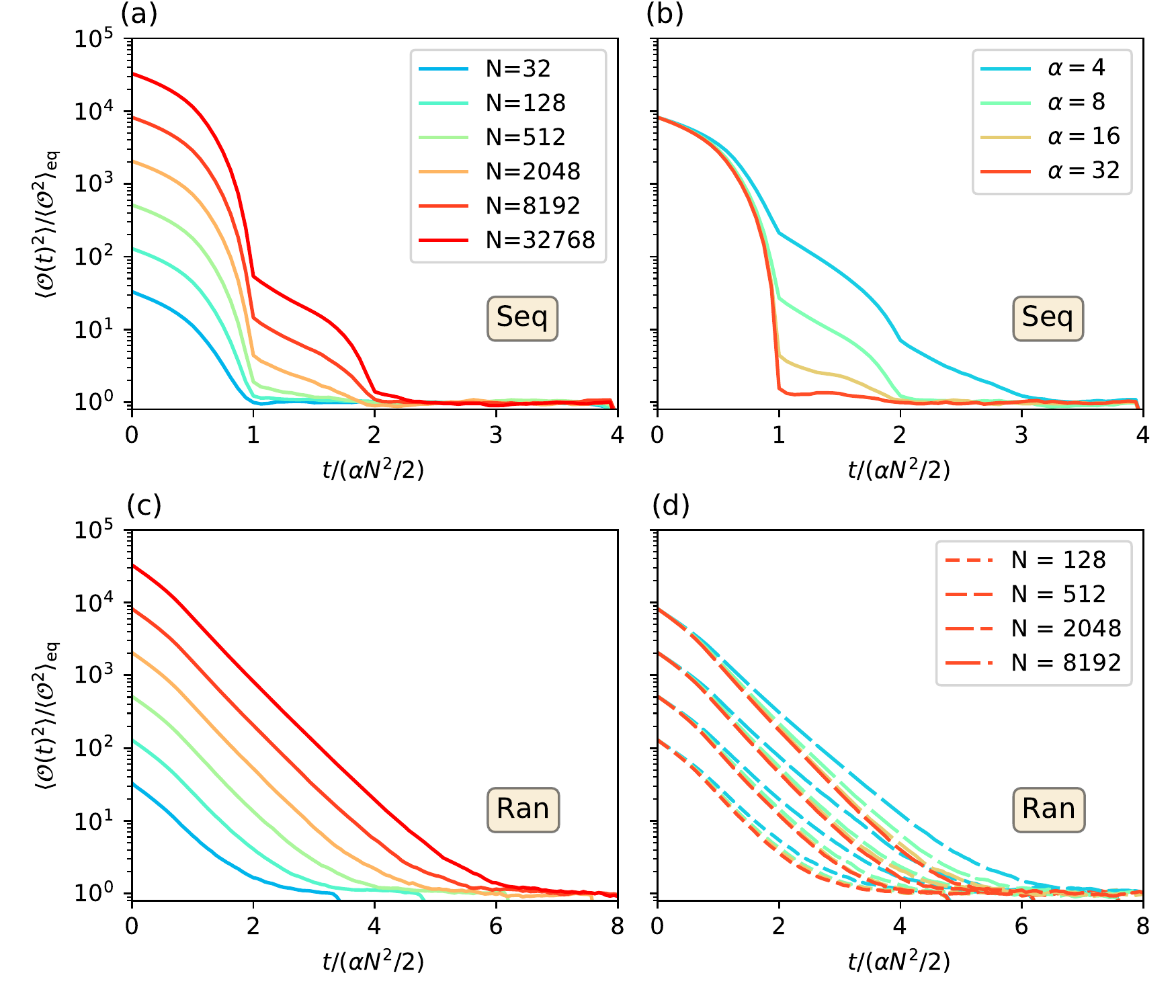}
\caption{Crossover from the discrete lifted algorithm to ECMC, \emph{via} the 
variance of the mid-system distance $x_{i+N/2} - x_i$ for ordered $x_i$, 
started from compact initial condition (see
\cite{KapferKrauth2017}). Discrete step size 
with $\epsilon \sim \unifb{0, \Lfree/N/\alpha}$, and chain length $l \sim 
\unifd{\alpha, \alpha N}$
\subcap{a}:
Sequential lifted Metropolis with
constant $\alpha=10$ for different $N$: 
The 
cross-over from perfect sampling for small $N$ at a time scale \bigO{N^2} 
towards \bigO{N^2 \log N} appears evident. 
\subcap{b}: 
Sequential algorithm for $N=8192$, with increasing $\alpha$: 
\bigO{N^2} mixing scale emerges for large $\alpha$.
\subcap{c}: 
Random lifted Metropolis  with 
$\alpha=10$ for different $N$ (legend as in \subcap{a}): 
\bigO{N^2 \log N} mixing time scale (conjectured 
earlier\cite{KapferKrauth2017}).
\subcap{d}: 
Random lifted forward Metropolis algorithm: Limited role of $\alpha$ 
(color code for $\alpha$ as in \subcap{b}).
}
\label{fig:CompactRelax}
\end{figure}

ECMC is the limit of the lifted forward Metropolis algorithm,
and the sequential ECMC the limit of the sequential lifted forward
Metropolis algorithm for step sizes much smaller than the mean free
space between spheres ($\mean{\epsilon} = \Lfree/(2N \alpha)$ with $\alpha \gg
1$). For a given discretization $2/\alpha$,  and for small $N$, 
the sequential lifted forward algorithm mimics the \bigO{N^2} mixing of 
the sequential ECMC, but 
for large $N$, it seems to cross over into \bigO{N^2 \log N} mixing
(see \subfig{fig:CompactRelax}{a}). \bigO{N^2} mixing also emerges at 
fixed $N$ for large 
$\alpha$ (see 
\subfig{fig:CompactRelax}{b}). 
(This is obtained using the heuristic mid-system variance $x_{i+ N/2} - x_i$ 
for ordered $x_i$, see \cite{KapferKrauth2017}.) 
In contrast, the random lifted forward 
Metropolis algorithm shows 
\bigO{N^2 \log N} mixing (see \subfig{fig:CompactRelax}{c}), 
as discussed earlier \cite{KapferKrauth2017}. This scaling
is little influenced by the discretization (see \subfig{fig:CompactRelax}{d}). 
It thus appears that the $N \to \infty $ 
limit of the sequential lifted forward Metropolis algorithm does not commute
with the small discretization limit $\alpha \to \infty$.

\section{Conclusions}
In this paper we have proven that for 1D hard spheres,  ECMC with a uniform
distribution of chain length $\ell \sim [0,\Lfree]$, with $\Lfree = L -
Nd$ realizes a perfect sample in \bigO{N^2 \log N} events that correspond
to \bigO{N \log N} chains. This confirms, in a special case, an earlier
conjecture\cite{KapferKrauth2017} for the mixing time of ECMC.  
For this case, we can compute the \TVD\ 
but also indicate a stopping rule for a time
(which depends on the particular realization of the Markov chain), after
which the Markov chain is in equilibrium. We have also provided numerical
evidence that the $N^2 \log N$ mixing prevails for other distributions of
$\ell$, namely for the uniform distribution $\unifb{0, \lambda \Lfree}$
and the Gaussian, and used the coupon-collector approximation to justify this 
approximation.

We have furthermore discussed a sequential ECMC which mixes in a time
\bigO{N^2}. For this algorithm, \quot{particle swaps} are essential. We have 
checked that the discrete version of this algorithm, namely the sequential 
lifted forward Metropolis algorithm crosses over, as the number $N$ of spheres 
is increased, to an \bigO{N^2 \log N} mixing behavior. In this formula, the 
origin of the 
logarithm is unclear, as it can no longer stem from the coupon 
collector. It would be of great interest for the fundamental 
understanding of irreversible MCMC algorithms to extend the results from ECMC to 
discrete versions, that is the lifted forward Metropolis algorithm and its 
sequential variant, as well as to the corresponding lattice models that may be 
easier to treat.

The lessons from our analysis of 1D hard-sphere systems are threefold. First, 
irreversible Markov chains can be proven to mix on shorter
time scales than reversible algorithms. Second, the speed of ECMC depends on the
whole distribution of the chain lengths $\ell$, but not on its mean value. 
Third, sequential-update algorithms (that remain valid in higher
dimensions) can mix on faster time scales than random-update versions. 
It remains to be seen how these lessons carry over to more intricate models and 
to higher dimensions.

\begin{acknowledgments}
We thank Florent Krzakala for a helpful discussion.
W.K. acknowledges support from the Alexander von Humboldt Foundation.
\end{acknowledgments}

% 
% \bibliographystyle{eplbib.bst}%abbrv}
% \bibliography{General.bib,TASEP.bib}\addcontentsline{toc}{section}{
% Literaturverzeichnis
%  }

\newpage

\renewcommand\refname{ }
\end{document}